\documentclass[%
prd
 ,twocolumn%
 ,secnumarabic%
,amssymb, amsmath,nobibnotes,showpacs]{revtex4}
\usepackage{bm}%
\usepackage{graphicx,calc,epsfig,pstricks,bbm}
\usepackage{tikz}
\expandafter\ifx\csname package@font\endcsname\relax\else
 \expandafter\expandafter
 \expandafter\usepackage
 \expandafter\expandafter
 \expandafter{\csname package@font\endcsname}%
\fi

\newcommand {\be}{\begin{equation}}
\newcommand {\ee}{\end{equation}}
\newcommand{\ba}{\begin{array}{c}}
\newcommand{\ea}{\end{array}}
\newcommand{\scr}{\scriptscriptstyle}
\newcommand{\cube}{\ba
 \begin{tikzpicture}
\pgfmathsetmacro{\cubex}{0.15}
\pgfmathsetmacro{\cubey}{0.15}
\pgfmathsetmacro{\cubez}{0.15}
\draw (0,0,0) -- ++(-\cubex,0,0) -- ++(0,-\cubey,0) -- ++(\cubex,0,0) -- cycle;
\draw (0,0,0) -- ++(0,0,-\cubez) -- ++(0,-\cubey,0) -- ++(0,0,\cubez) -- cycle;
\draw (0,0,0) -- ++(-\cubex,0,0) -- ++(0,0,-\cubez) -- ++(\cubex,0,0) -- cycle;
\end{tikzpicture}
\ea}

\begin{document}
\title{A new perspective on cosmology in Loop Quantum Gravity}%

\author{Emanuele \surname{Alesci}}
\email{emanuele.alesci@gravity.fau.de}
\affiliation{Institute for Quantum Gravity, FAU Erlangen-N\"urnberg, Staudtstr. 7, D-91058 Erlangen, Germany, EU}
\affiliation{Instytut Fizyki Teoretycznej, Uniwersytet Warszawski, ul. Ho{\.z}a 69, 00-681 Warszawa, Poland, EU}
\author{Francesco \surname{Cianfrani}}
\email{francesco.cianfrani@gravity.fau.de}
\affiliation{Institute for Quantum Gravity, FAU Erlangen-N\"urnberg, Staudtstr. 7, D-91058 Erlangen, Germany, EU}


\date{\today}%

\begin{abstract}
We present a new cosmological model derived from Loop Quantum Gravity. The formulation is based on a projection of the kinematical Hilbert space of the full theory down to a subspace representing the proper arena for an inhomogeneous Bianchi I model. This procedure gives a direct link between the full theory and its cosmological sector. The emerging quantum cosmological model represents a simplified arena on which the complete canonical quantization program can be tested. The achievements of this analysis could also shed light on Loop Quantum Cosmology and its relation with the full theory.  
\end{abstract}


\maketitle

\section*{} 

{\it Introduction-} A proper quantum description for the gravitational field is expected to avoid the emergence of those singularities which plague General Relativity (GR). In particular, a Quantum Gravity (QG) scenario should tame the initial singularity proper of a homogeneous and eventually isotropic space-time, which does not permit to follow the backward evolution of our Universe before the Big Bang. Therefore, the implementation of a quantum scenario in such symmetry-reduced models gives a privileged arena where the viability of the proposed QG model can be tested. 

This analysis is particularly useful in the Loop Quantum Gravity (LQG) \cite{revloop}. This framework provide us with a consistent scheme for the canonical quantization of geometric degrees of freedom. However, it is difficult to extract predictions from LQG, particularly due to the complicated expression of the super-Hamiltonian operator \cite{qsd,ar} based on the volume operator \cite{vol1,vol2} for which no close analytical formulas are available. 

In symmetry-reduced models the dynamical problem is simplified, thus also the quantum description is expected to be easier with respect to the full theory. Hence, the implementation of LQG in a cosmological space-time could offer a unique opportunity to test the viability of this fundamental description of the gravitational field. 

A well settled cosmological scenario derived from LQG is described by the theory now called Loop Quantum Cosmology (LQC) \cite{revlqc1,revlqc2}. In LQC, {\it one first reduces the phase-space} according with the symmetries of a homogeneous space-time and {\it then quantize} using LQG techniques. However it has not yet been shown that the cosmological sector of LQG is LQC. Moreover the quantization of classically reduced variables restricts the regime of applicability of the theory to isotropic or anisotropic cosmologies but there is not space left for inhomogeneities in the theory (apart from perturbation theory \cite{pertLQC}).
A natural question is then: What happens the other way around, in which one {\it first quantizes} GR, obtaining LQG, and {\it then reduces} at a quantum level according to the desired symmetries? An answer to this question would definitely set up LQC from LQG, 
and it could also open the road to the study of quantum inhomogeneous cosmological models.

{\it This letter presents a framework in which this answer can be given.}

The main result of LQC is the replacement of the initial singularity with a bounce occurring when the energy density of the clock-like scalar field is comparable with a critical energy density \cite{ashprl}. This critical energy density is related with the existence of a discrete polymer-like structure in minisuperspace and its value is fixed by the minimum nonvanishing eigenvalue of the area operator in LQG \cite{vol1}. 
The emergence of such a scale is a necessary tool to define a proper action for the super-Hamiltonian operator in LQC. But, this feature marks a close difference with the full theory, where the same scale plays the role of a regulator and it is removed in the regularization procedure. The reason for this tension between LQG and LQC can be traced back to diffeomorphisms invariance, which in the former is implemented by defining states over s-knots \cite{sknots} on which the super-Hamiltonian can be regularized and the Dirac algebra quantized without anomalies \cite{revloop}, while in the latter it is lost ab-initio by the choice of variables. 

Therefore, in order to define a quantum formulation for a cosmological space-time which is closer to LQG than LQC, it should be defined a proper reduction from the full theory which preserves a kind of reduced diffeomorphisms invariance  such that a regularized expression for the super-Hamiltonian can be defined. This is done in this paper by considering an inhomogeneous extension of the Bianchi I cosmological model. This extension can be relevant both for the late Universe evolution, as a reparametrization of the homogeneous phase, and in the early stages, as providing an effective description for each Kasner epoch during the Mixmaster regime \cite{BKL}. 


Hence, in this letter we discuss the projection of the full theory down to a reduced model describing the inhomogeneous Bianchi I space-time (for the details see the companion paper \cite{comp}). The invariance under the full diffeomorphisms group is broken down to a proper subgroup (reduced diffeomorphisms) and the fundamental structure of the quantum space-time is that of a cubical cell. A key-point of this projection is $SU(2)$ gauge invariance: a gauge-fixing of the $SU(2)$ group is realized by mimicking the analogous procedure adopted in Spin-Foam models to solve the so-called simplicity constraints \cite{sfsimpl}. This procedure selects the quantum states of the reduced model out of the gauge invariant functionals of $SU(2)$ group elements. It turns out that some nontrivial intertwiners map the representations associated to different edges. The recoupling theory of the $SU(2)$ group induces a reduced recoupling theory in the cosmological model. The reduced volume operator is diagonal and it is possible to evaluate explicitly the action of the un-regularized super-Hamiltonian operator on a generic element of the reduced Hilbert space. Moreover, one can construct the s-knots states associated with reduced diffeomorphisms, over which the super-Hamiltonian can be regularized. 

Therefore, the quantum dynamics of the proposed cosmological model is well defined. The emerging picture is that of a cubical lattice, whose edges are endowed with $U(1)$ quantum numbers determining the geometrical properties. The reduced SU(2) intertwiners at the nodes allow us to realize quantum configurations with space-dependent quantum numbers, thus they can accommodate the inhomogeneous content of the theory.

{\it Loop Quantum Gravity-} LQG is based on a reformulation of gravity in which configuration variables 
are the Ashtekar-Barbero connections $A^i_a$, and the canonically conjugate momenta are the inverse densitized triads $E^a_i$ associated with the spatial metric. In this approach, besides the standard invariance under 3-diffeomorphisms generated by the vector constraint $V_a$ and time reparametrizations generated by the Hamiltonian constraint $H$, there is a further SU(2) gauge symmetry generated by the Gauss constraint $\mathcal{G}$. The whole phase space can be parametrized by considering $SU(2)$ holonomies $h_e$ and the fluxes $E_i(S)$ defined over all the possible paths $e$ and surfaces $S$ of the spatial manifold. These coordinates are particularly useful: $SU(2)$ gauge transformations map $h_e$ in $h_e'=\lambda_{s(e)}h_e\lambda^{-1}_{t(e)}$, $s(e)$ and $t(e)$ being the initial and final points of $e$, respectively, while $\lambda$ denotes $SU(2)$ group elements. The action of finite diffeomorphisms $\varphi$ maps the original holonomy into the one evaluated on the transformed path, $h_e\rightarrow h_{\varphi(e)}$. Hence, the quantization is realized applying Dirac program \cite{ht} by representing the holonomy-flux algebra in a kinematical Hilbert space $\mathcal{H}^{kin}$ and solving the three constraints $\mathcal{G}=0, V_a=0, H=0$ on it. 
$\mathcal{H}^{kin}$ is composed by the cylindrical functions \cite{revloop} whose basis elements are parametrized by all the possible graphs $\Gamma$ and they are defined as the tensor product of Wigner matrices  $D^{j_e}(h_e)$ in the $SU(2)$ irreducible representations $j_e$ for each edge, {\it i.e.}
\be
\langle h|\Gamma, j_e,\rangle=\bigotimes_{e\in\Gamma} D^{j_e}(h_e).
\label{spinnetworks}
\ee

The action of fluxes $E_i(S)$ across a surface $S$ is nonvanishing only if there is a non-empty intersection between $e$ and $S$, such that $S$ splits $e$ in $e_1\bigcup e_2$, and it is given by 
\be
E_i(S)D^{(j_e)}(h_e)
=8\pi\gamma l_P^2 \; o(e,S) \; D^{j_e}(h_{e_1})\,^{(j_e)}\tau^{i}\,D^{j_e}(h_{e_2}).\label{Eop}
\ee  

$\gamma$ and $l_P$ being the Immirzi parameter and the Planck length, while the factor $o(e,S)=0,1,-1$ according to the relative sign of $e$ and the normal to $S$. $^{(j_e)}\tau_i$ denotes the SU(2) generator in $j_e$-dimensional representation.  

The operator generating local $SU(2)$ gauge transformations $\lambda(x)$ acts on the basis elements (\ref{spinnetworks}) as follows
\be
U_{\mathcal{ G}}(g) D^j_{mn}(h_e)= D^j_{mn}(\lambda_{s(e)}h_e \lambda^{-1}_{t(e)}).
\ee
The Gauss constraint is implemented by a group averaging procedure defining 
a projector $P_{\mathcal{ G}}$ to the $SU(2)$-invariant Hilbert space $^{\mathcal{G}}\mathcal{H}^{kin}$, by integrating over the $SU(2)$ group elements $\lambda_{s(e)}$ and $\lambda_{t(e)}$ for each edge.  As a result the basis elements (\ref{spinnetworks}) are replaced by
 \be
< h |\Gamma,\{j_e\},\{x_v\}>=\prod_{v\in\Gamma} \prod_{e\in\Gamma} {x_{v}}\cdot D^{j_{e}}(h_{e}),  
\label{spinnet}
\ee

$x_{v}$ being the $SU(2)$ invariant intertwiners between the representations associated with the edges emanating from $v$ and $\cdot$ means index contraction. States of the form \eqref{spinnet} are called spinnetworks.
 
The diffeomorphisms constraint $V_a$ is solved by group averaging spinnetworks over the equivalence class of knots defining the so called s-knots \cite{sknots}.

The superhamiltonian operator $H$ can be written as a discrete sum by adopting a graph-dependent triangulation and by replacing connections with holonomies along some additional edges $s$. The resulting expression converges to the classical limit as the triangulation gets finer and finer, which means that the lattice spacing, as well as the length of the edges $s$, vanishes. This limit is well-defined in the diffeomorphisms invariant Hilbert space \cite{qsd,ar}. However, the matrix elements of H are very complicated \cite{ioantonia} and there is not closed analytical form for them.

{\it Inhomogeneous Bianchi I model-} Let us assume the following form for phase-space variables
\be
E^a_i=p_i(t,x)\delta^a_i,\qquad A^i_a(t,x)= c_i(t,x)\delta^i_a.\label{varr}
\ee

These conditions can be obtained exactly by considering a re-parametrized Bianchi I model, in which each scale factor contains a dependence on the associated Cartesian coordinate $x^i$.  The same result holds approximatively in the case of the generalized Kasner solution \cite{gkas}, in which one deals with an inhomogeneous extension of a Bianchi I model and the spatial gradients of the scale factors can be neglected with respect to time-derivatives. This solution describes the behavior of the early Mixmaster Universe during each Kasner epoch \cite{BKL}. 

The SU(2) gauge-fixing associated with the conditions (\ref{varr}) can be written as \cite{io2}
\be
\chi_i=\epsilon_{ij}^{\phantom{12}k}E^a_k\delta_a^j=0.\label{gcon}
\ee

The conditions (\ref{varr}) are implemented in the kinematical Hilbert space of LQG i)by considering only edges $e_i$ parallel to fiducial vectors $\omega_i=\partial_i$ and ii)by the restriction from $SU(2)$ to $U(1)$ group elements via the imposition of the condition \eqref{gcon}. 

The requirement i) can be realized via a projectors $P$ which acts on holonomies $h_e$ such that $Ph_e=h_e$ if $e=e_i$ for some $i$, otherwise it vanishes. Let us call $\mathcal{H}_{P}$ the resulting Hilbert space,  we can project down to $\mathcal{H}_{P}$  the action $U(\varphi_\xi)$ of a generic diffeomorphisms $\varphi_\xi$, $\xi$ being the infinitesimal parameters, as follows
\be
{}^{red}\!U(\varphi_\xi)=PU(\varphi_\xi)P.
\ee

It turns out that ${}^{red}\!U$ is nonvanishing and connected to the identity only if 
\be
\xi^i=\xi^i(x^i).
\ee

Each $\xi^i$ is then the infinitesimal parameter of an arbitrary translation along the direction $i$ and a rigid translation along other directions. Therefore, only a subgroup of the whole 3-diffeomorphiosms group is represented in $\mathcal{H}_P$ and we call its elements reduced diffeomorphisms.

As soon the condition (\ref{gcon}) is concerned, it cannot be implemented as first-class condition, because it is a gauge-fixing and it does not commute with the SU(2) Gauss constraint. Henceforth, we implement Eq. (\ref{gcon}) weakly, by mimicking the procedure adopted in Spin-Foam models to impose the simplicity constraints \cite{sfsimpl}. 

Building holonomies that fulfill the second condition into Eq. (\ref{varr}) one obtains the reduced holonomies $\tilde{h}_{e_i}$. These on a classical level are elements of the $U(1)_i$ subgroups ($i=1,2,3$) obtained by stabilizing the $SU(2)$ group around the internal directions $(1,0,0)$, $(0,1,0)$ and $(0,0,1)$, respectively.  Now how can we consistently implement eq. \eqref{varr} in the kinematical Hilbert space of LQG ${\mathcal H}^{kin}$ ? The condition on the $A^i_a$ has to be imposed on the $SU(2)$ holonomies and it amounts to the requirement that the $SU(2) $ cylindrical functions are completely determined by their restrictions to the $U(1)_i$ subgroups; to this aim we lift the functions $\psi(h_i)$ of $U(1)_i$ group elements $h_i$ to those of the $SU(2)$ elements $g$ via the Dupuis-Livine map \cite{dl}:  
\be
\tilde{\psi}(g)=\int_{U(1)} dh_i\; K(g,h_i) \psi(h_i), \qquad g\in SU(2),
\label{projected1}
\ee
with the Kernel $K(g,h_i)$ given by 
\be
K(g,h_i)=\sum_{n} \int_{U(1)_i}  dk_i\; \chi^{j(n)}(gk_i) \chi^n(k_ih_i),
\ee
$\chi^{j(n)}(g)$ and $\chi^n(h_i)$ being the $SU(2)$ and $U(1)_i$ characters, respectively. By expanding $\psi$ in the Peter-Weyl base for $U(1)_i$ with coefficients $\psi^{n_i}$, one gets
\be
\begin{split}
\tilde{\psi}(g)_{e_i}&=\sum_{n_i} {}^i\!D^{j(n_i)}_{m=n_i\,r=n_i
}(g) \psi^{n_i}_{e_i},
\end{split}
\label{projected 3}
\ee
${}^i\!D^{j(n_i)}_{mr}$ being the Wigner matrices in the spin base $|j,m\rangle_i$ that diagonalize the operators $J^2$ and $J_i$.
 The condition on $E^a_i$ can then be imposed in the quantum theory, considering the fluxes across the surfaces $S^i$ dual to $e_i$ and imposing the associated condition \eqref{gcon} via a master constraint $\chi^2=\sum_i(\chi_i)^2=0$.  This way, one finds an approximate solution to  $\chi^2 \tilde{\psi}(g)_{e_i}=0$ in the limit $j\rightarrow+\infty$ by fixing $j(n_i)=|n_i|$. This selects the degree of the representation $j(n_i)$ ({\it i.e} the relation between the $SU(2)$ and the $U(1)_i$ quantum numbers).

Therefore, the proper quantum states adapted to the restrictions (\ref{varr}) read
\be
\tilde{\psi}(g)_{e_i}=\sum_{n_i} {}^i\!D^{j=|n_i|}_{n_i\,n_i}(g) \psi^{n_i}_{e_i}=\sum_{j} {}^i\!D^{|j|}_{jj}(g) \psi^{j}_{e_i}.
\label{fine projected}
\ee

One can then define a new projector $P_\chi$ from $\mathcal{H}_{P}$ to a subspace $\mathcal{H}^{R}$ whose elements are of the form (\ref{fine projected}). In $\!\mathcal{H}^R$ the following relations hold weakly:
\be
E_k(S^l)\tilde{\psi}(g)_{e_i}=8\pi\gamma  l_P^2 \delta_k^l\delta_i^k\sum_j j\, {}^i\!D^j_{jj}(g)\psi^j_{e_i}.\label{flussi sulle ridotte}
\ee 

Therefore, the only nonvanishing fluxes are $E_i(S^i)$ and the resulting holonomy-flux algebra is the same as in reduced quantization \cite{comp}. However, because we have established a projection from $\mathcal{H}^{kin}$ to $\mathcal{H}^R$, we can discuss the implications of the original $SU(2)$ gauge invariance. In this respect,   
let us consider the projector $P_{\mathcal{G}}$ to $^{\mathcal{G}}\mathcal{H}^{kin}$ and let us project it down to the reduced Hilbert space. This will give us a new projector $P_{\mathcal{G},\chi}=P^{\dag}_{\chi}\; P_{\mathcal{G}} P_{\chi}$ acting into $\mathcal{H}^R$, which selects the gauge invariant reduced Hilbert space $^{\mathcal{G}}\mathcal{H}^{R}$ whose basis states are
\be
\langle h|\Gamma, j_e, x_v\rangle_R= \prod_{v\in\Gamma} \prod_{e\in\Gamma} \langle {\bf j_i}, {\bf x}|{\bf j_i},  \vec{{\bf e}}_i \rangle \cdot \;{}^i\!D^{j_{e_i}}_{j_{e_i} j_{e_i}}(h_{e_i}).
\ee

The coefficients $\langle{\bf j_i}, {\bf x}|{\bf j_i},  \vec{{\bf e}}_i \rangle$ are the projection of Livine-Speziale coherent intertwiners $|{\bf j_i},  \vec{{\bf e}}_i \rangle$ \cite{ls}  (with normals $\vec{{\bf e}}_i$ tangent to the cubical cell) on the usual intetwiner base $|{\bf j_i}, {\bf x}\rangle$ and they are the intertwiners of the reduced model (reduced intertwiners).

It is possible to give the following diagrammatic representation elucidating the relation between reduced elements ${}^i\!D^{|j|}_{jj}(g)$ and $SU(2)$ group elements  
\be
{}^i\!D^{|j|}_{jj}(g)=\begin{array} {c} 
\ifx\JPicScale\undefined\def\JPicScale{1.1}\fi
\psset{unit=\JPicScale mm}
\psset{linewidth=0.3,dotsep=1,hatchwidth=0.3,hatchsep=1.5,shadowsize=1,dimen=middle}
\psset{dotsize=0.7 2.5,dotscale=1 1,fillcolor=black}
\psset{arrowsize=1 2,arrowlength=1,arrowinset=0.25,tbarsize=0.7 5,bracketlength=0.15,rbracketlength=0.15}
\begin{pspicture}(15,0)(68,8)
\psline{|*-}(52,3)(48,3)
\rput(35,8){$\scriptstyle{j}$}
\rput{0}(35,3){\psellipse[](0,0)(3,-3)}
\rput(35,3){$g$}
\pspolygon[](42,6)(48,6)(48,0)(42,0)
\rput(45,3){$R_{\vec{e_i}}$}
\psline[fillstyle=solid](38,3)(42,3)
\psline{}(32,3)(28,3)
\pspolygon[](22,6)(28,6)(28,0)(22,0)
\rput(25,3){$R^{-1}_{\vec{e_i}}$}
\psline[fillstyle=solid]{|-}(18,3)(22,3)
\end{pspicture}
\end{array}
\label{h ridotta}
\ee
\\
where the solid lines represent the identity in the base $| j,m\rangle$ that diagonalizes $\tau^3$, while the projection on the maximum magnetic number is given by $|j,j\rangle=
\begin {array} {c}
\ifx\JPicScale\undefined\def\JPicScale{1}\fi
\psset{unit=\JPicScale mm}
\psset{linewidth=0.3,dotsep=1,hatchwidth=0.3,hatchsep=1.5,shadowsize=1,dimen=middle}
\psset{dotsize=0.7 2.5,dotscale=1 1,fillcolor=black}
\psset{arrowsize=1 2,arrowlength=1,arrowinset=0.25,tbarsize=0.7 5,bracketlength=0.15,rbracketlength=0.15}
\begin{pspicture}(0,0)(10,8)
\psline[fillstyle=solid]{-|}(2,3)(7,3)
\rput(4,2){$\scriptstyle{j}$}
\end{pspicture}
\end{array}$. The group elements are represented by boxes or circles depending if they denote the Wigner matrix of specific fixed rotation $R_{\vec{e}_i}$ that move the $e_3$ axis to the $e_i$ axis, selecting the desired $U(1)_i$ subgroup, or a generic $SU(2)$ Wigner matrix $D(g)$ in the $|j,m\rangle$ base, respectively. For instance, in this notation a 3-valent reduced intertwiner can be represented as follows
\be
 \langle j_1,j_2, j_3 , x_v | j_1,j_2, j_3 , e_1, e_2, e_3 \rangle=\begin{array} {c}
\ifx\JPicScale\undefined\def\JPicScale{1}\fi
\psset{unit=\JPicScale mm}
\psset{linewidth=0.3,dotsep=1,hatchwidth=0.3,hatchsep=1.5,shadowsize=1,dimen=middle}
\psset{dotsize=0.7 2.5,dotscale=1 1,fillcolor=black}
\psset{arrowsize=1 2,arrowlength=1,arrowinset=0.25,tbarsize=0.7 5,bracketlength=0.15,rbracketlength=0.15}
\begin{pspicture}(35,100)(75,120)
\psline[fillstyle=solid]{-|}(42.83,101.83)(40,99)
\pspolygon[](46,103)(44,101)(42,103)(44,105)
\psline[fillstyle=solid](51,110)(44.84,103.84)
\rput(40,104){$\scr{R_{e_1}}$}
\rput(43,108){$\scr{j_1}$}
\psline[fillstyle=solid]{-|}(51,110)(51,120)
\psline[fillstyle=solid]{-|}(59.17,101.83)(62,99)
\pspolygon[](58,105)(60,103)(58,101)(56,103)
\rput(62,104){$\scr{R_{e_2}}$}
\rput(57,107){$\scr{j_2}$}
\psline[fillstyle=solid](51,110)(57.16,103.84)
\rput(48,115){$\scr{j_3}$}
\end{pspicture}
\end{array}
\label{3-valent reduced}
\ee

This diagrammatic notation allows us to represent arbitrary elements of $^{\mathcal{G}}\mathcal{H}^{R}$ by combining  reduced holonomies \eqref{h ridotta} and reduced intertwiners like \eqref{3-valent reduced}. The rules of SU(2) recoupling theory induce then a reduced recoupling by which the action of any operator can be computed.
 
{\it Volume-} Given a region $\Omega$ containing for simplicity only one vertex $v$ of the graph $\Gamma$, a regularized expression for the volume can be defined as in the full theory \cite{vol1,vol2} in terms of the fluxes $E_i(S^j)$, where $S^1\cap S^2\cap S^3=v$. Furthermore, each $E_i(S^j)$ is nonvanishing only if $i=j$ and it acts only on holonomies $h_i$. As a consequence, the volume operator is diagonal in $^{\mathcal{G}}\mathcal{H}^{R}$ and the action of the super-Hamiltonian operator can be explicitly evaluated. 

{\it SuperHamiltonian-} Let us consider the Euclidean part of the super-Hamiltonian in the full theory regularized via a cubulation of the spatial manifold adapted to the cubical graph $\Gamma$. This can be done at each node $v$ by choosing for each pair of links $e_i$ and $e_j$ incident at $v$ a semi-analytic arcs $a_{ij}$ that respect the lattice structure, i.e. such that the end points $s_{e_i},s_{e_j}$ are interior points of $e_i,e_j$, respectively, and  $a_{ij}\cap\Gamma=\{s_{e_i},s_{e_j}\}$. The arc $s_i$ is the segment of $e_i$($e_j$) from $v$ to $s_i$($s_j$), while $s_{i}$, $s_{j}$ and $a_{ij}$ generate a rectangle $\alpha_{ij} := s_{i} \circ a_{ij} \circ s_j^{-1}$. By repeating this construction for all $e_i$ and $e_j$ at $v$ and for all $v\in\Gamma$, one gets a complete cubulation $C$ of the spatial manifold. The quantum super-Hamiltonian can be constructed {\it a-la} Thiemann \cite{qsd}
and projected down to 
$^{\mathcal{G}}\mathcal{H}^{R}$
by replacing the volume operator and holonomies of the full theory with the corresponding reduced expressions, ${}^{R}V$ and $\widetilde{h}$, respectively. This way one finds the reduced super-Hamiltonian ${}^{R}H^m_{\cube}$
\be
{}^{R}H^m_{\cube}[N]:= \frac{N(v)}{N^2_m} \,  \, \epsilon^{ijk} \,
   \mathrm{Tr}\Big[\tilde{h}^{(m)}_{\alpha_{ij}} \tilde{h}^{(m)}_{s_{k}} \big\{\tilde{h}^{(m)-1}_{s_{k}},{}^{R}V\big\}\Big], 
   \label{hred}
\ee
 where the trace and the reduced holonomies are in an arbitrary irreducible representation $m$, while the factor $N_m$ is a normalizing factor. The expression above converges to the classical value as the cubulation gets finer and finer. 
The regulator of the super-Hamiltonian (\ref{hred}) is given by the lattice spacing of the cubulation, which can be changed via a reduced diffeomorphisms. As a consequence, on the reduced-diffeomorphisms invariant class of knots the regulator can be safely removed.

{\it Conclusions-}
We discussed a procedure to derive a quantum formulation for a Bianchi I space-time directly from the full LQG framework. This procedure is based on a restriction on the set of admissible graphs, such that only those with a cubical vertex structure are selected, and on the implementation of the SU(2) gauge fixing via a projection to the group elements stabilizing some internal directions. As a result we inferred a reduced kinematical Hilbert space whose edges are labeled by $U(1)$ representation and some nontrivial intertwiners are predicted at vertexes. The presence of a residual diffeomrphisms invariance allowed us to construct reduced s-knots over which the super-Hamiltonian is well defined and the quantization is anomaly free as in the full theory. 
This scheme offers an alternative point of view on the cosmological sector of LQG. Hence, the analysis of the dynamical implications may sustain the LQC paradigm and the bouncing scenario, elucidating the origin of the polymer parameter. 
It is worth noting how such an analysis is going to be addressed in forthcoming investigations, thanks to the simplifications occurring in the reduced kinematical Hilbert space (in particular the diagonal form of the volume operator) which will allow us to compute explicitly the action of the operator (\ref{hred}) and the solutions of the quantum Einstein equations .
Moreover, this formulation can also be applied to an inhomogeneous Bianchi I model under the assumptions proper of the Belinski-Lipschitz-Kalatnikov conjecture \cite{BKL}. Therefore, the proposed scheme can characterise the quantum behavior of the early inhomogeneous Universe during each Kasner epoch, giving us some indications on the fate of the generic cosmological solution in LQG. 

{\it Acknowledgment}   
The authors wish to thank T.Thiemann and K.Giesel for useful discussions.
F.C. thanks ``Angelo Della Riccia'' foundation for its financial support. 
This work was partially supported by the grant of Polish Narodowe Centrum
Nauki nr 2011/02/A/ST2/00300.
This work has been partially realized in the framework of the CGW collaboration (www.cgwcollaboration.it).

\end{document}